\documentclass[aps,prb,reprint,superscriptaddress,nobibnotes]{revtex4-1}

\usepackage{amsmath} 
\usepackage{amsfonts} 
\usepackage{graphicx} 

\begin{document}

\title{Scanning gate microscopy of ultra clean carbon nanotube quantum dots}

\author{Jiamin Xue}
\altaffiliation[Present address: ]{School of Physical Sciences and Technologies, ShanghaiTech University, Shanghai 201203, China}
\affiliation{Physics Department, University of Arizona, 1118 E 4th Street, Tucson, AZ 85721, USA}
\author{Rohan Dhall}
\author{Stephen B. Cronin}
\affiliation{Department of Electrical Engineering, University of Southern California, Los Angeles, California 90089, USA}
\author{Brian J. LeRoy}
\email{leroy@physics.arizona.edu}
\affiliation{Physics Department, University of Arizona, 1118 E 4th Street, Tucson, AZ 85721, USA}

\date{\today}

\begin{abstract}
We perform scanning gate microscopy on individual suspended carbon nanotube quantum dots.  The size and position of the quantum dots can be visually identified from the concentric high conductance rings. For the ultra clean devices used in this study, two new effects are clearly identified. Electrostatic screening creates non-overlapping multiple sets of Coulomb rings from a single quantum dot.  In double quantum dots, by changing the tip voltage, the interactions between the quantum dots can be tuned from the weak to strong coupling regime.
\end{abstract}

\maketitle

\section{Introduction} 

Quantum dots have been proposed as one of the building blocks in future quantum information processing schemes~\cite{Loss1998}. Carbon nanotube quantum dots are especially attractive because they are expected to have a long spin relaxation time~\cite{Bulaev2008}, which makes them preferable for quantum bits using the electron spin. Understanding and controlling quantum dot behavior is crucial for potential applications. To this end, electrical transport measurements have been widely used to probe the energy levels of single and coupled quantum dots~\cite{Kouwenhoven1997,Jarillo2004,Mason2004,Steele2009}. However, this type of measurement lacks information on the real space structure of the quantum dot system. Scanning tunneling spectroscopy was also used to provide atomic resolution images of the wavefunctions in quantum dots~\cite{Lemay2001} but special techniques are needed to simultaneously perform electrical transport measurements~\cite{LeRoy2007}.  Scanning gate microscopy (SGM) is a more versatile technique for probing quantum dots in carbon nanotubes because it does not require a conducting substrate~\cite{Woodside2002,Bockrath2001,Tans2000,Kalinin2002,Freitag2002}. 

SGM is a powerful technique for probing quantum systems because it only requires that the tip is capacitively coupled to the system being probed.  It does not require a tunnel current or physical contact.  Therefore the method has been used to probe a wide range of systems from semiconductor two-dimensional electron gases~\cite{Topinka2000} to graphene quantum dots~\cite{Jalilian2011,Schnez2011}. In SGM, the charged tip acts as movable gate and the conductance through the device as a function of tip position is recorded.  As the tip scans over the nanotube, it perturbs the local band structure causing the energy levels on the quantum dot to change.  The changing energy levels on the nanotube lead to a varying number of electrons and hence changes in the conductance.  The resulting two-dimensional conductance map shows the location and size of the quantum dot structures.

Previous SGM measurements of carbon nanotubes used devices fabricated on SiO$_2$ surfaces, which had many charged impurities, which resulted in a complicated electrostatic environment~\cite{Woodside2002,Bockrath2001,Tans2000,Kalinin2002,Freitag2002}. As the tip scanned over the sample, it not only affected the nanotube, but also modified charge states of nearby impurities. Hence, the SGM maps represented a mixture of intrinsic properties of the nanotube quantum dots and complicated perturbations from the environment.  To overcome this limitation, we performed SGM measurements of suspended ultra clean carbon nanotubes~\cite{LeRoy2004,Cao2005}. From the SGM maps, we can clearly see the location and size of single and multiple quantum dots, without complications due to the substrate. By changing the voltage applied to the tip, we can also modify the coupling strength between the double quantum dots.

\section{Device Fabrication}

Figure 1(a) shows a schematic diagram of the experimental setup. A trench of ~600 nm deep and 0.5 to 1.5 $\mu$m wide is predefined in a Si/SiO$_2$/Si$_3$N$_4$ wafer. Source, drain, and gate electrodes are written with photo-lithography followed by deposition of Pt metal as electrodes~\cite{Bushmaker2009a,Bushmaker2009b}.  Small windows, 1.5 $\mu$m by 1.5 $\mu$m, are patterned in photoresist near the edges of the trench and ferric nitrate is dispersed as the catalyst for nanotube growth.  Carbon nanotubes are grown by chemical vapor deposition using a mixture of argon gas bubbled through ethanol and hydrogen at 825 $^\circ$C.  This is performed as the final step of sample preparation to avoid surface contamination~\cite{LeRoy2004,Cao2005}.  Some of the nanotubes will grow across the trench forming a working device. We use electrical transport measurements at room temperature to identify conductive devices and Raman spectroscopy to select devices with only a single carbon nanotube. The sample is transferred to an ultrahigh vacuum system and cooled down to 4 K for the SGM measurements.  In this paper, we report measurements on two typical types of devices, the first is a single quantum dot and the second is a controllable double quantum dot. 

\begin{figure}[ht]
\includegraphics[width=8.5cm]{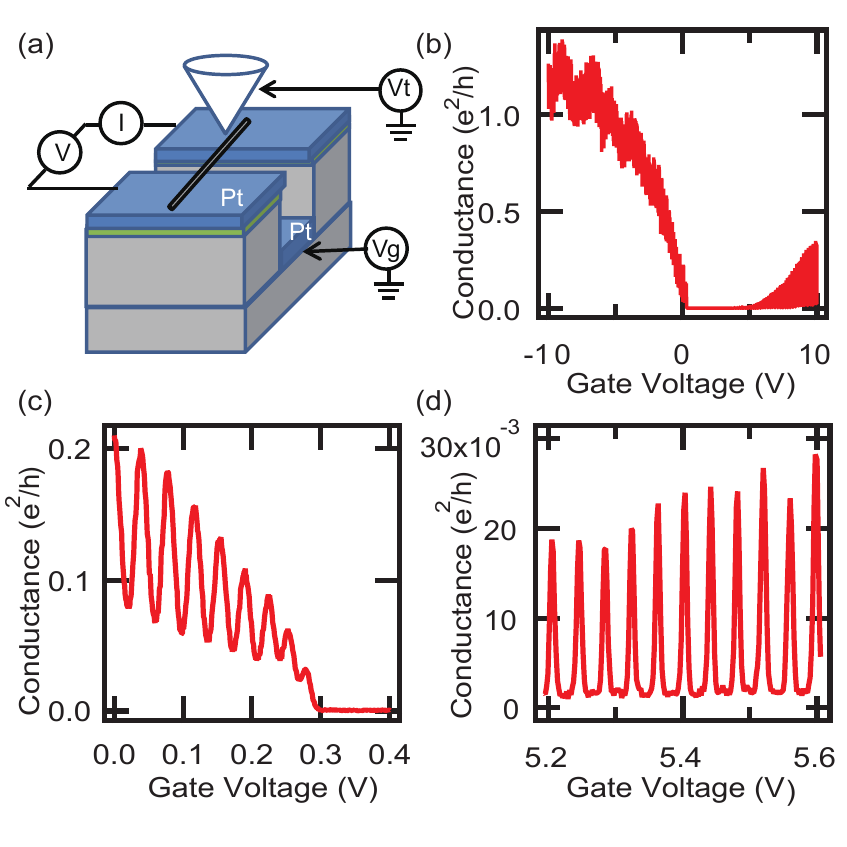} 
\caption{Scanning gate microscopy setup and electrical transport measurements. (a) Schematic of the experimental setup showing the electrical connections to the nanotube and the tip. (b) Conductance of a nanotube as a function of gate voltage at 4 K. (c) and (d) zoom in to the hole side and electron side of (b), respectively. }
\label{fig:schematic}
\end{figure}

\section{Experimental Results}

Figure 1(b) shows an electrical transport measurement of a single quantum dot device at a temperature of 4 K. The conductance through the nanotube is measured by applying an AC voltage of 0.4 mV to the nanotube and measuring the resulting current using lock-in detection.  This device has a 500 nm wide trench. Due to hole doping from the Pt electrodes, the upper edge of the nanotube valence band is lined up with the Fermi level of the electrodes~\cite{Tans1998}. This can be seen from Fig. 1(b), where the onset of conduction through the valence band starts near zero gate voltage, while transport through the conduction band does not start until a gate voltage of nearly +5V.  Looking at the electrical transport through the two different bands, we observe very different behavior.  For the hole side (Fig. 1(c)), we observe Fabry-Perot type oscillations~\cite{Liang2001} with a high average conductance.  This is a sign of very transparent contacts between the valence band of the nanotube and the Pt electrodes.  In contrast on the electron side (Fig. 1(d)), very regular Coulomb blockade peaks can be seen, indicating much higher tunneling barriers to the conduction band.  The regularity of both the Fabry-Perot oscillations and the Coulomb blockade peaks implies that the nanotube is free of defects and behaves as a single quantum dot.

Figures 2(a)-(d) show spatial maps of the conductance through the nanotube plotted as a function of tip position. Bright color means high conductance and dark color the opposite. In order to obtain these images, the nanotube is held at ground with a small AC voltage (0.4 mV) applied across it. The tip with a fixed voltage applied to it is raster scanned 200 nm above the nanotube. The conductance through the nanotube at each tip position is recorded with a lock-in amplifier and displayed as the two dimensional images shown in Figs. 2(a)-(d).  All of these images were acquired with a gate voltage of +8 V, i.e. in the Coulomb blockade regime on the electron side.  However, the tip voltages varied from -1.5 V in Fig. 2(a) to +2 V in Fig. 2(d).  The position of the nanotube is indicated by the solid horizontal line in Fig. 2(a), and the edges of the electrodes are shown by the two vertical dashed lines. The distance between the electrodes is 500 nm. A set of concentric rings resulting from a single quantum dot is clearly seen in Fig. 2(a), with the tip voltage at -1.5 V. The rings indicate positions where the number of electrons on the nanotube quantum dot changes by one.  The tip induces charge on the nanotube due to its voltage and the capacitance between it and the nanotube.  Therefore, since the tip voltage is held constant, the rings represent locations of constant induced charge or capacitance.  When the tip moves from left to right along the center line approaching the position of the nanotube, an electron is removed from the quantum dot each time it crosses a ring.  After passing the center of the rings, electrons are added to the quantum dot each time the tip passes over a ring. 

\begin{figure*}
\includegraphics[width=17.8cm]{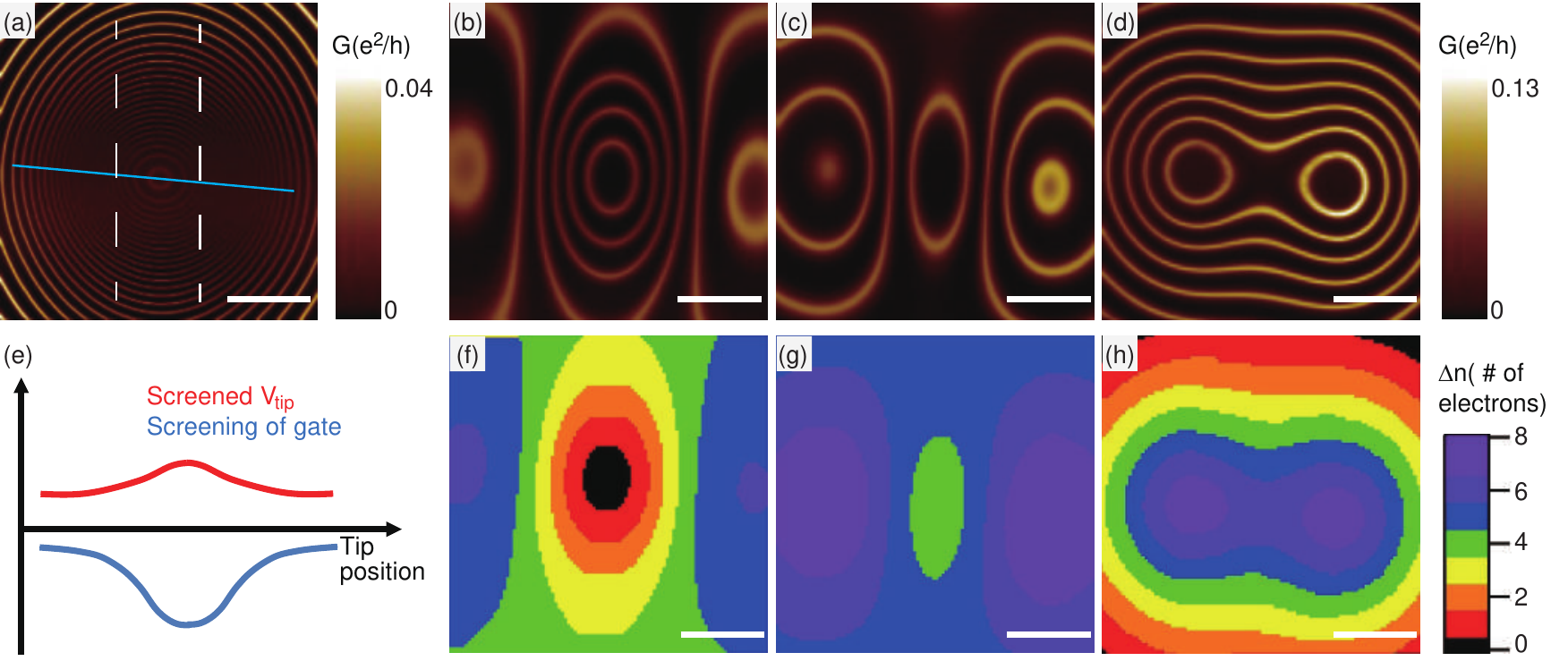} 
\caption{SGM images of transport through a single quantum dot and maps of induced charge. (a) to (d): SGM maps of a single quantum dot device. All images were taken at the same gate voltage (+8 V) but with different tip voltages. The tip voltages are as follows (a) -1.5 V, (b) +0.8 V, (c) +1.2 V and (d) +2 V. Dashed lines in (a) represent the edges of the trench. The solid line indicates the nanotube. (e) Qualitative evolution of the screening effect of the gate by the tip (blue curve) and tip by the electrodes (red curve), as the positively charged tip scans along the nanotube. (f) to (h): Maps of induced charge on the nanotube corresponding to (b) to (d). Numbers on the color bar only have a relative meaning (i.e. areas in the map with green color have one more electron than areas with yellow color). All scale bars are 500 nm. }
\label{fig:images}
\end{figure*}

When the tip voltage is increased from -1.5 V in Fig. 2(a) to +0.8 V in Fig. 2(b), the SGM image changes completely.  The circularly symmetric rings in Fig. 2(a) shrink and become elliptical at the center of the image while two new side rings emerge. When the tip voltage is further increased to +1.2 V in Fig. 2(c) and +2 V in Fig. 2(d), the side rings grow and the center rings eventually disappear. To understand the evolution of behavior from Figs. 2(a)-(d), we need to find the induced charge on the nanotube as a function of tip voltage and position.  The tip has two effects; (1) it induces charge on the nanotube due to its capacitive coupling. (2) It screens the effect of the underlying gate electrode. Due to screening of the gate by the tip, even with zero tip voltage, the nanotube behaves as if a voltage that is opposite to the gate voltage is applied to the tip. In the case of Fig. 2 (+8 V gate voltage), zero voltage on the tip is effectively a negative tip voltage for the nanotube. The presence of the large source and drain electrodes reduces the screening effect of the tip when it is over the electrodes. So the tip is effectively more negative in the center of the trench than above the electrodes (see Fig. 2(e), blue curve). In Fig. 2(a), the tip voltage is -1.5 V, so the tip always removes electrons from the nanotube. However, when the tip voltage is positive [Figs. 2(b)-(d)], the applied positive voltage competes with the effective negative voltage due to the screening of the gate (see Fig. 2(e), red curve).  This makes the tip positive above the electrodes and negative above the trench. So as the tip moves closer to the center of the side rings, more electrons are added onto the nanotube; on the other hand, as the tip moves closer to the center of the middle ring, more electrons are removed from the nanotube.  Maps of induced charge on the nanotube shown in Figs. 2(f)-(h), corresponding to Figs. 2(b)-(d), clearly show the opposite behavior of the side and middle rings.  There are always fewer electrons in the center ring compared to the side rings.

Now that the effect of the tip is understood for a single quantum dot system, we examine how it can be used to control the coupling in a double quantum dot device.  Figure 3(a) shows an electrical transport measurement at 4K of a device with a 1.5 $\mu$m wide trench. Compared with the device in Fig. 1(b), it has lower conductance and Coulomb blockade peaks appear on both sides of the band gap, implying very opaque tunnel barriers at the contacts for both electrons and holes. The Coulomb blockade peaks are not regularly spaced as in Fig. 1(d), which implies the nanotube consists of more than one quantum dot in series.  The SGM image in Fig. 3(c) (gate voltage is -2 V and tip voltage is 1 V) clearly confirms this conclusion as it can be viewed as two sets of rings superimposed on each other, where each set originates from a single quantum dot. Since the two quantum dots are in series, there is non-zero conductance only at points when the two sets of rings intersect, where both of the dots are conductive. This results in the pattern seen in Fig. 3(c) showing isolated points of conductance. When the tip voltage decreases to -2.4 V [Fig. 3(d)], the SGM pattern evolves from isolated points to a series of avoided crossings. This transition occurs when the coupling between the two quantum dots increases from weak to strong~\cite{Steele2009,Livermore1996}. The tip is now used to not only image the quantum dots but also control the coupling between them. 

\begin{figure}[ht]
\includegraphics[width=8.5cm]{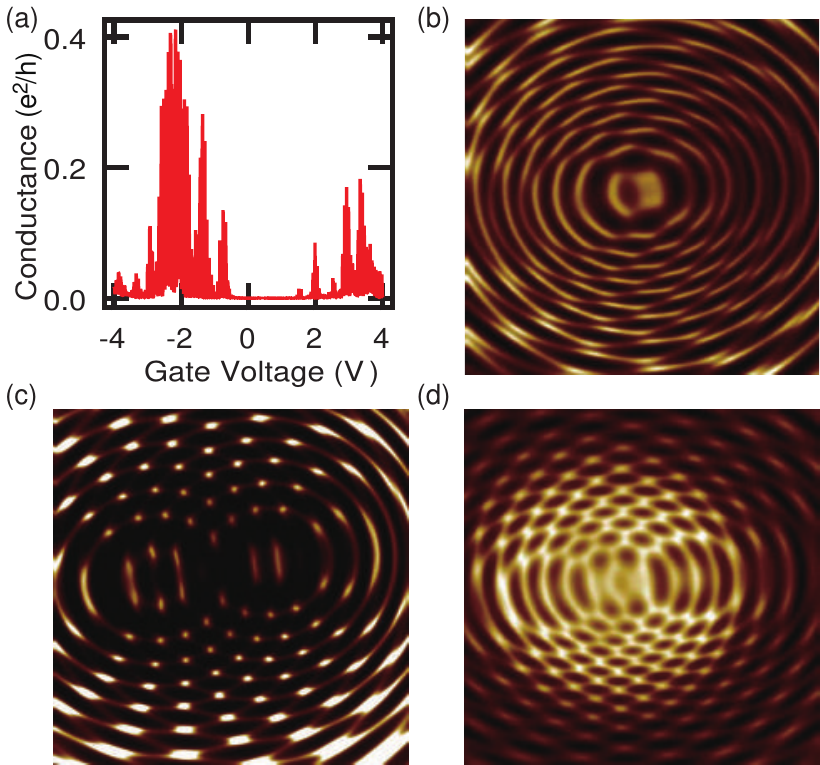} 
\caption{Electrical transport and SGM measurements of a double quantum dot.  (a) Conductance of double quantum dot device as a function of gate voltage at 4 K. (b) SGM map of the device with gate voltage at 3 V and tip voltage at -1 V. (c) SGM map with gate voltage at -2 V and tip voltage at 1 V. (d) SGM map with gate voltage at -2 V and tip voltage at  -2.4 V. Scale bars in all images are 500 nm.}
\label{fig:double}
\end{figure}

We can understand how the tip controls the double dot coupling from the band diagrams shown in Figs. 4(a) and (c), corresponding to the weakly and strongly coupled case respectively. In Fig. 4(a), gray dashed lines represent the edges of the valence and conduction bands with zero gate voltage applied and no tip present.  At the contacts the Fermi level is pinned to the valence band while near the center of the trench the nanotube is undoped.  The green dashed lines represent the effect on the bands when -2 V is applied to the gate. Its effect is strongest for the center portion of the nanotube, and gradually decreases to zero near the contacts due to pinning and screening by them. Now the tip with +1 V applied is brought above the nanotube. As explained in the first part of this paper, the applied voltage on the tip and screening of the gate by the tip work together since they are both positive in the case of Fig. 4(a), so the tip is positive as seen by the nanotube. A positive tip will push down the bands as shown by the red dashed lines, which represent the effect of the tip. Combining the two effects (green dashed lines and red dashed lines) together, the final band edges are shown as the blue solid lines.  Near the valence band edge, there are two quantum dots, which are weakly coupled because of the large barrier in the middle.  

We can simulate the SGM image of a weakly coupled double dot system (shown in Fig. 4(b)). In the simulation, two quantum dots are capacitively coupled, with a small interdot capacitance representing weak coupling~\cite{Livermore1996}. Conductance through the double dot depends exponentially on the energy difference between the source-drain Fermi energy and the energy of the next available electron level of each dot. So the conductance is exponentially suppressed except when both dots have their next available electron level lined up with the source-drain Fermi energy. This implies that the double quantum dot only conducts when tip induces a changing number of electrons on each dot, i.e. when the conductance rings intersect.  The simulated SGM image in Fig. 4(b) agrees well with the experimental SGM image in Fig. 3(c).  On the other hand, when the tip voltage is -2.4 V, the behavior changes.  In this case, the applied negative tip voltage exceeds the effective positive voltage resulting from screening of the gate, so the net tip voltage is negative as seen by the nanotube. This works together with the negative gate voltage to push up the bands resulting in a strongly coupled double dot system. In the simulation, the interdot capacitance is set high to represent the stronger inter dot coupling, giving rise to a series of avoided crossings as shown in Fig. 4(d), which agrees well with the experimental image in Fig. 3(d). So far, we have only discussed the behavior when negative gate voltage is applied to the device. When positive gate voltage is applied and the bands of the nanotube are pulled down enough, the left and right quantum dots start to couple through the conduction band and Klein tunneling~\cite{Steele2009,Katsnelson2006} turns them into a well-coupled double quantum dot system. This can be seen from the SGM image in Fig. 3(b) with +3 V gate voltage. 

\begin{figure}[ht]
\includegraphics[width=8.5cm]{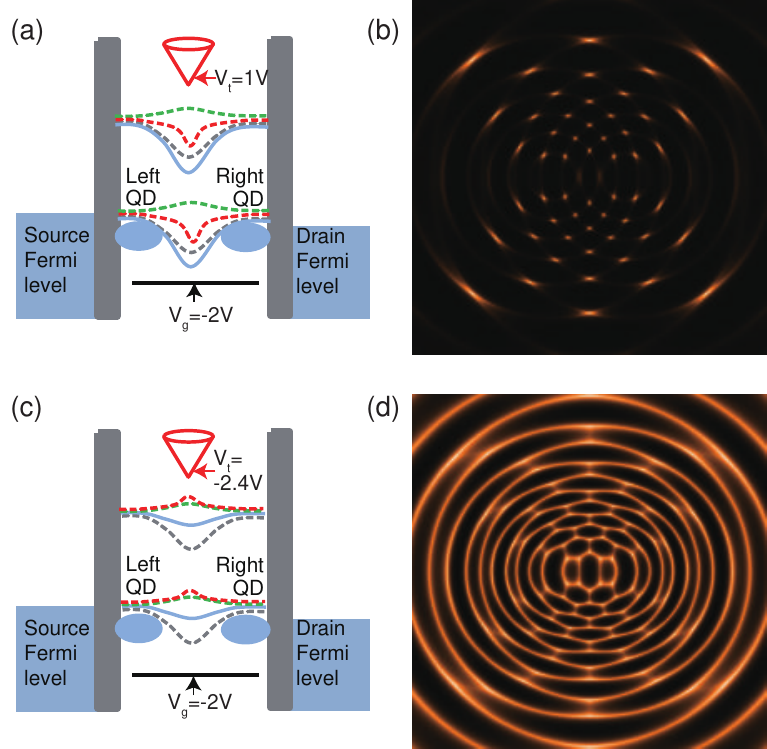} 
\caption{Explanation of double quantum dot control and simulated SGM images. (a) Band diagram for tip-induced weakly coupled double quantum dot. Gray dashed lines represent the band edges with zero gate voltage and no tip present; green dashed lines represent the effect on the bands when -2 V is applied to the gate; red dashed lines represent the effect of a +1 V tip; blue solid lines are final band edges with these two effects taken into account. (b) Simulated SGM map for a weakly coupled double quantum dot. (c) Band diagram for tip induced strongly coupled double quantum dot. The different lines have similar meaning as in (a). Notice the red dashed lines are now upward due to a negative tip voltage (-2.4 V). (d) Simulated SGM map for a strongly coupled double quantum dot. Scale bars are 500 nm.}
\label{fig:theory}
\end{figure}

\section{Conclusion}

In conclusion, we have presented scanning gate microscopy images of ultra clean nanotube quantum dots. We demonstrated the ability to observe quantum dot structure in real space and also the ability to control the coupling strength between multiple quantum dots in series. Future work should be done to bring the effect of the tip to the weak perturbation regime where electron interaction effects can be probed with the techniques presented in this paper.  By reducing the spatial extant of the tip perturbation, the nature of the wavefunctions inside the quantum dots should be able to be probed in these ultra clean nanotubes.

\begin{acknowledgments}

The authors thank Charles Stafford for helpful discussions. J.X. and B.J.L. acknowledge the support of NSF CAREER grant No. DMR/0953784. R.D. and S.C. acknowledge the support of ONR grant No. N000141010511 and DOE grant No. DE-FG02-07ER46376. A portion of this work was done in the UCSB nanofabrication facility, part of the NSF funded NNIN network.

\end{acknowledgments}

\bibliography{mybib}

\begin{thebibliography}{24}%
\makeatletter
\providecommand \@ifxundefined [1]{%
 \@ifx{#1\undefined}
}%
\providecommand \@ifnum [1]{%
 \ifnum #1\expandafter \@firstoftwo
 \else \expandafter \@secondoftwo
 \fi
}%
\providecommand \@ifx [1]{%
 \ifx #1\expandafter \@firstoftwo
 \else \expandafter \@secondoftwo
 \fi
}%
\providecommand \natexlab [1]{#1}%
\providecommand \enquote  [1]{``#1''}%
\providecommand \bibnamefont  [1]{#1}%
\providecommand \bibfnamefont [1]{#1}%
\providecommand \citenamefont [1]{#1}%
\providecommand \href@noop [0]{\@secondoftwo}%
\providecommand \href [0]{\begingroup \@sanitize@url \@href}%
\providecommand \@href[1]{\@@startlink{#1}\@@href}%
\providecommand \@@href[1]{\endgroup#1\@@endlink}%
\providecommand \@sanitize@url [0]{\catcode `\\12\catcode `\$12\catcode
  `\&12\catcode `\#12\catcode `\^12\catcode `\_12\catcode `\%12\relax}%
\providecommand \@@startlink[1]{}%
\providecommand \@@endlink[0]{}%
\providecommand \url  [0]{\begingroup\@sanitize@url \@url }%
\providecommand \@url [1]{\endgroup\@href {#1}{\urlprefix }}%
\providecommand \urlprefix  [0]{URL }%
\providecommand \Eprint [0]{\href }%
\providecommand \doibase [0]{http://dx.doi.org/}%
\providecommand \selectlanguage [0]{\@gobble}%
\providecommand \bibinfo  [0]{\@secondoftwo}%
\providecommand \bibfield  [0]{\@secondoftwo}%
\providecommand \translation [1]{[#1]}%
\providecommand \BibitemOpen [0]{}%
\providecommand \bibitemStop [0]{}%
\providecommand \bibitemNoStop [0]{.\EOS\space}%
\providecommand \EOS [0]{\spacefactor3000\relax}%
\providecommand \BibitemShut  [1]{\csname bibitem#1\endcsname}%
\let\auto@bib@innerbib\@empty
\bibitem [{\citenamefont {Loss}\ and\ \citenamefont
  {DiVincenzo}(1998)}]{Loss1998}%
  \BibitemOpen
  \bibfield  {author} {\bibinfo {author} {\bibfnamefont {D.}~\bibnamefont
  {Loss}}\ and\ \bibinfo {author} {\bibfnamefont {D.~P.}\ \bibnamefont
  {DiVincenzo}},\ }\href@noop {} {\bibfield  {journal} {\bibinfo  {journal}
  {Physical Review A}\ }\textbf {\bibinfo {volume} {57}},\ \bibinfo {pages}
  {120} (\bibinfo {year} {1998})}\BibitemShut {NoStop}%
\bibitem [{\citenamefont {Bulaev}\ \emph {et~al.}(2008)\citenamefont {Bulaev},
  \citenamefont {Trauzettel},\ and\ \citenamefont {Loss}}]{Bulaev2008}%
  \BibitemOpen
  \bibfield  {author} {\bibinfo {author} {\bibfnamefont {D.~V.}\ \bibnamefont
  {Bulaev}}, \bibinfo {author} {\bibfnamefont {B.}~\bibnamefont {Trauzettel}},
  \ and\ \bibinfo {author} {\bibfnamefont {D.}~\bibnamefont {Loss}},\
  }\href@noop {} {\bibfield  {journal} {\bibinfo  {journal} {Physical Review
  B}\ }\textbf {\bibinfo {volume} {77}},\ \bibinfo {pages} {235301} (\bibinfo
  {year} {2008})}\BibitemShut {NoStop}%
\bibitem [{\citenamefont {Sohn}\ \emph {et~al.}(1997)\citenamefont {Sohn},
  \citenamefont {Kouwenhoven} \emph {et~al.}}]{Kouwenhoven1997}%
  \BibitemOpen
  \bibfield  {author} {\bibinfo {author} {\bibfnamefont {L.~L.}\ \bibnamefont
  {Sohn}}, \bibinfo {author} {\bibfnamefont {L.~P.}\ \bibnamefont
  {Kouwenhoven}},  \emph {et~al.},\ }\href@noop {} {\emph {\bibinfo {title}
  {Mesoscopic Electron Transport: Proceedings of the NATO Advanced Study
  Institute, 25 June-5 July 1996, Curacao, Netherlands Antilles}}},\ Vol.\
  \bibinfo {volume} {345}\ (\bibinfo  {publisher} {Springer Science \& Business
  Media},\ \bibinfo {year} {1997})\BibitemShut {NoStop}%
\bibitem [{\citenamefont {Jarillo-Herrero}\ \emph {et~al.}(2004)\citenamefont
  {Jarillo-Herrero}, \citenamefont {Sapmaz}, \citenamefont {Dekker},
  \citenamefont {Kouwenhoven},\ and\ \citenamefont {van~der
  Zant}}]{Jarillo2004}%
  \BibitemOpen
  \bibfield  {author} {\bibinfo {author} {\bibfnamefont {P.}~\bibnamefont
  {Jarillo-Herrero}}, \bibinfo {author} {\bibfnamefont {S.}~\bibnamefont
  {Sapmaz}}, \bibinfo {author} {\bibfnamefont {C.}~\bibnamefont {Dekker}},
  \bibinfo {author} {\bibfnamefont {L.~P.}\ \bibnamefont {Kouwenhoven}}, \ and\
  \bibinfo {author} {\bibfnamefont {H.~S.}\ \bibnamefont {van~der Zant}},\
  }\href@noop {} {\bibfield  {journal} {\bibinfo  {journal} {Nature}\ }\textbf
  {\bibinfo {volume} {429}},\ \bibinfo {pages} {389} (\bibinfo {year}
  {2004})}\BibitemShut {NoStop}%
\bibitem [{\citenamefont {Mason}\ \emph {et~al.}(2004)\citenamefont {Mason},
  \citenamefont {Biercuk},\ and\ \citenamefont {Marcus}}]{Mason2004}%
  \BibitemOpen
  \bibfield  {author} {\bibinfo {author} {\bibfnamefont {N.}~\bibnamefont
  {Mason}}, \bibinfo {author} {\bibfnamefont {M.}~\bibnamefont {Biercuk}}, \
  and\ \bibinfo {author} {\bibfnamefont {C.}~\bibnamefont {Marcus}},\
  }\href@noop {} {\bibfield  {journal} {\bibinfo  {journal} {Science}\ }\textbf
  {\bibinfo {volume} {303}},\ \bibinfo {pages} {655} (\bibinfo {year}
  {2004})}\BibitemShut {NoStop}%
\bibitem [{\citenamefont {Steele}\ \emph {et~al.}(2009)\citenamefont {Steele},
  \citenamefont {Gotz},\ and\ \citenamefont {Kouwenhoven}}]{Steele2009}%
  \BibitemOpen
  \bibfield  {author} {\bibinfo {author} {\bibfnamefont {G.}~\bibnamefont
  {Steele}}, \bibinfo {author} {\bibfnamefont {G.}~\bibnamefont {Gotz}}, \ and\
  \bibinfo {author} {\bibfnamefont {L.}~\bibnamefont {Kouwenhoven}},\
  }\href@noop {} {\bibfield  {journal} {\bibinfo  {journal} {Nature
  nanotechnology}\ }\textbf {\bibinfo {volume} {4}},\ \bibinfo {pages} {363}
  (\bibinfo {year} {2009})}\BibitemShut {NoStop}%
\bibitem [{\citenamefont {Lemay}\ \emph {et~al.}(2001)\citenamefont {Lemay},
  \citenamefont {Janssen}, \citenamefont {van~den Hout}, \citenamefont {Mooij},
  \citenamefont {Bronikowski}, \citenamefont {Willis}, \citenamefont {Smalley},
  \citenamefont {Kouwenhoven},\ and\ \citenamefont {Dekker}}]{Lemay2001}%
  \BibitemOpen
  \bibfield  {author} {\bibinfo {author} {\bibfnamefont {S.~G.}\ \bibnamefont
  {Lemay}}, \bibinfo {author} {\bibfnamefont {J.~W.}\ \bibnamefont {Janssen}},
  \bibinfo {author} {\bibfnamefont {M.}~\bibnamefont {van~den Hout}}, \bibinfo
  {author} {\bibfnamefont {M.}~\bibnamefont {Mooij}}, \bibinfo {author}
  {\bibfnamefont {M.~J.}\ \bibnamefont {Bronikowski}}, \bibinfo {author}
  {\bibfnamefont {P.~A.}\ \bibnamefont {Willis}}, \bibinfo {author}
  {\bibfnamefont {R.~E.}\ \bibnamefont {Smalley}}, \bibinfo {author}
  {\bibfnamefont {L.~P.}\ \bibnamefont {Kouwenhoven}}, \ and\ \bibinfo {author}
  {\bibfnamefont {C.}~\bibnamefont {Dekker}},\ }\href@noop {} {\bibfield
  {journal} {\bibinfo  {journal} {Nature}\ }\textbf {\bibinfo {volume} {412}},\
  \bibinfo {pages} {617} (\bibinfo {year} {2001})}\BibitemShut {NoStop}%
\bibitem [{\citenamefont {LeRoy}\ \emph {et~al.}(2007)\citenamefont {LeRoy},
  \citenamefont {Heller}, \citenamefont {Pahilwani}, \citenamefont {Dekker},\
  and\ \citenamefont {Lemay}}]{LeRoy2007}%
  \BibitemOpen
  \bibfield  {author} {\bibinfo {author} {\bibfnamefont {B.~J.}\ \bibnamefont
  {LeRoy}}, \bibinfo {author} {\bibfnamefont {I.}~\bibnamefont {Heller}},
  \bibinfo {author} {\bibfnamefont {V.~K.}\ \bibnamefont {Pahilwani}}, \bibinfo
  {author} {\bibfnamefont {C.}~\bibnamefont {Dekker}}, \ and\ \bibinfo {author}
  {\bibfnamefont {S.~G.}\ \bibnamefont {Lemay}},\ }\href@noop {} {\bibfield
  {journal} {\bibinfo  {journal} {Nano letters}\ }\textbf {\bibinfo {volume}
  {7}},\ \bibinfo {pages} {2937} (\bibinfo {year} {2007})}\BibitemShut
  {NoStop}%
\bibitem [{\citenamefont {Woodside}\ and\ \citenamefont
  {McEuen}(2002)}]{Woodside2002}%
  \BibitemOpen
  \bibfield  {author} {\bibinfo {author} {\bibfnamefont {M.~T.}\ \bibnamefont
  {Woodside}}\ and\ \bibinfo {author} {\bibfnamefont {P.~L.}\ \bibnamefont
  {McEuen}},\ }\href@noop {} {\bibfield  {journal} {\bibinfo  {journal}
  {Science}\ }\textbf {\bibinfo {volume} {296}},\ \bibinfo {pages} {1098}
  (\bibinfo {year} {2002})}\BibitemShut {NoStop}%
\bibitem [{\citenamefont {Bockrath}\ \emph {et~al.}(2001)\citenamefont
  {Bockrath}, \citenamefont {Liang}, \citenamefont {Bozovic}, \citenamefont
  {Hafner}, \citenamefont {Lieber}, \citenamefont {Tinkham},\ and\
  \citenamefont {Park}}]{Bockrath2001}%
  \BibitemOpen
  \bibfield  {author} {\bibinfo {author} {\bibfnamefont {M.}~\bibnamefont
  {Bockrath}}, \bibinfo {author} {\bibfnamefont {W.}~\bibnamefont {Liang}},
  \bibinfo {author} {\bibfnamefont {D.}~\bibnamefont {Bozovic}}, \bibinfo
  {author} {\bibfnamefont {J.~H.}\ \bibnamefont {Hafner}}, \bibinfo {author}
  {\bibfnamefont {C.~M.}\ \bibnamefont {Lieber}}, \bibinfo {author}
  {\bibfnamefont {M.}~\bibnamefont {Tinkham}}, \ and\ \bibinfo {author}
  {\bibfnamefont {H.}~\bibnamefont {Park}},\ }\href@noop {} {\bibfield
  {journal} {\bibinfo  {journal} {Science}\ }\textbf {\bibinfo {volume}
  {291}},\ \bibinfo {pages} {283} (\bibinfo {year} {2001})}\BibitemShut
  {NoStop}%
\bibitem [{\citenamefont {Tans}\ and\ \citenamefont {Dekker}(2000)}]{Tans2000}%
  \BibitemOpen
  \bibfield  {author} {\bibinfo {author} {\bibfnamefont {S.~J.}\ \bibnamefont
  {Tans}}\ and\ \bibinfo {author} {\bibfnamefont {C.}~\bibnamefont {Dekker}},\
  }\href@noop {} {\bibfield  {journal} {\bibinfo  {journal} {Nature}\ }\textbf
  {\bibinfo {volume} {404}},\ \bibinfo {pages} {834} (\bibinfo {year}
  {2000})}\BibitemShut {NoStop}%
\bibitem [{\citenamefont {Kalinin}\ \emph {et~al.}(2002)\citenamefont
  {Kalinin}, \citenamefont {Bonnell}, \citenamefont {Freitag},\ and\
  \citenamefont {Johnson}}]{Kalinin2002}%
  \BibitemOpen
  \bibfield  {author} {\bibinfo {author} {\bibfnamefont {S.~V.}\ \bibnamefont
  {Kalinin}}, \bibinfo {author} {\bibfnamefont {D.~A.}\ \bibnamefont
  {Bonnell}}, \bibinfo {author} {\bibfnamefont {M.}~\bibnamefont {Freitag}}, \
  and\ \bibinfo {author} {\bibfnamefont {A.}~\bibnamefont {Johnson}},\
  }\href@noop {} {\bibfield  {journal} {\bibinfo  {journal} {Applied physics
  letters}\ }\textbf {\bibinfo {volume} {81}},\ \bibinfo {pages} {5219}
  (\bibinfo {year} {2002})}\BibitemShut {NoStop}%
\bibitem [{\citenamefont {Freitag}\ \emph {et~al.}(2002)\citenamefont
  {Freitag}, \citenamefont {Johnson}, \citenamefont {Kalinin},\ and\
  \citenamefont {Bonnell}}]{Freitag2002}%
  \BibitemOpen
  \bibfield  {author} {\bibinfo {author} {\bibfnamefont {M.}~\bibnamefont
  {Freitag}}, \bibinfo {author} {\bibfnamefont {A.}~\bibnamefont {Johnson}},
  \bibinfo {author} {\bibfnamefont {S.~V.}\ \bibnamefont {Kalinin}}, \ and\
  \bibinfo {author} {\bibfnamefont {D.~A.}\ \bibnamefont {Bonnell}},\
  }\href@noop {} {\bibfield  {journal} {\bibinfo  {journal} {Physical review
  letters}\ }\textbf {\bibinfo {volume} {89}},\ \bibinfo {pages} {216801}
  (\bibinfo {year} {2002})}\BibitemShut {NoStop}%
\bibitem [{\citenamefont {Topinka}\ \emph {et~al.}(2000)\citenamefont
  {Topinka}, \citenamefont {LeRoy}, \citenamefont {Shaw}, \citenamefont
  {Heller}, \citenamefont {Westervelt}, \citenamefont {Maranowski},\ and\
  \citenamefont {Gossard}}]{Topinka2000}%
  \BibitemOpen
  \bibfield  {author} {\bibinfo {author} {\bibfnamefont {M.}~\bibnamefont
  {Topinka}}, \bibinfo {author} {\bibfnamefont {B.}~\bibnamefont {LeRoy}},
  \bibinfo {author} {\bibfnamefont {S.}~\bibnamefont {Shaw}}, \bibinfo {author}
  {\bibfnamefont {E.}~\bibnamefont {Heller}}, \bibinfo {author} {\bibfnamefont
  {R.}~\bibnamefont {Westervelt}}, \bibinfo {author} {\bibfnamefont
  {K.}~\bibnamefont {Maranowski}}, \ and\ \bibinfo {author} {\bibfnamefont
  {A.}~\bibnamefont {Gossard}},\ }\href@noop {} {\bibfield  {journal} {\bibinfo
   {journal} {Science}\ }\textbf {\bibinfo {volume} {289}},\ \bibinfo {pages}
  {2323} (\bibinfo {year} {2000})}\BibitemShut {NoStop}%
\bibitem [{\citenamefont {Jalilian}\ \emph {et~al.}(2011)\citenamefont
  {Jalilian}, \citenamefont {Jauregui}, \citenamefont {Lopez}, \citenamefont
  {Tian}, \citenamefont {Roecker}, \citenamefont {Yazdanpanah}, \citenamefont
  {Cohn}, \citenamefont {Jovanovic},\ and\ \citenamefont
  {Chen}}]{Jalilian2011}%
  \BibitemOpen
  \bibfield  {author} {\bibinfo {author} {\bibfnamefont {R.}~\bibnamefont
  {Jalilian}}, \bibinfo {author} {\bibfnamefont {L.~A.}\ \bibnamefont
  {Jauregui}}, \bibinfo {author} {\bibfnamefont {G.}~\bibnamefont {Lopez}},
  \bibinfo {author} {\bibfnamefont {J.}~\bibnamefont {Tian}}, \bibinfo {author}
  {\bibfnamefont {C.}~\bibnamefont {Roecker}}, \bibinfo {author} {\bibfnamefont
  {M.~M.}\ \bibnamefont {Yazdanpanah}}, \bibinfo {author} {\bibfnamefont
  {R.~W.}\ \bibnamefont {Cohn}}, \bibinfo {author} {\bibfnamefont
  {I.}~\bibnamefont {Jovanovic}}, \ and\ \bibinfo {author} {\bibfnamefont
  {Y.~P.}\ \bibnamefont {Chen}},\ }\href@noop {} {\bibfield  {journal}
  {\bibinfo  {journal} {Nanotechnology}\ }\textbf {\bibinfo {volume} {22}},\
  \bibinfo {pages} {295705} (\bibinfo {year} {2011})}\BibitemShut {NoStop}%
\bibitem [{\citenamefont {Schnez}\ \emph {et~al.}(2011)\citenamefont {Schnez},
  \citenamefont {G{\"u}ttinger}, \citenamefont {Stampfer}, \citenamefont
  {Ensslin},\ and\ \citenamefont {Ihn}}]{Schnez2011}%
  \BibitemOpen
  \bibfield  {author} {\bibinfo {author} {\bibfnamefont {S.}~\bibnamefont
  {Schnez}}, \bibinfo {author} {\bibfnamefont {J.}~\bibnamefont
  {G{\"u}ttinger}}, \bibinfo {author} {\bibfnamefont {C.}~\bibnamefont
  {Stampfer}}, \bibinfo {author} {\bibfnamefont {K.}~\bibnamefont {Ensslin}}, \
  and\ \bibinfo {author} {\bibfnamefont {T.}~\bibnamefont {Ihn}},\ }\href@noop
  {} {\bibfield  {journal} {\bibinfo  {journal} {New Journal of Physics}\
  }\textbf {\bibinfo {volume} {13}},\ \bibinfo {pages} {053013} (\bibinfo
  {year} {2011})}\BibitemShut {NoStop}%
\bibitem [{\citenamefont {LeRoy}\ \emph {et~al.}(2004)\citenamefont {LeRoy},
  \citenamefont {Lemay}, \citenamefont {Kong},\ and\ \citenamefont
  {Dekker}}]{LeRoy2004}%
  \BibitemOpen
  \bibfield  {author} {\bibinfo {author} {\bibfnamefont {B.}~\bibnamefont
  {LeRoy}}, \bibinfo {author} {\bibfnamefont {S.}~\bibnamefont {Lemay}},
  \bibinfo {author} {\bibfnamefont {J.}~\bibnamefont {Kong}}, \ and\ \bibinfo
  {author} {\bibfnamefont {C.}~\bibnamefont {Dekker}},\ }\href@noop {}
  {\bibfield  {journal} {\bibinfo  {journal} {Applied physics letters}\
  }\textbf {\bibinfo {volume} {84}},\ \bibinfo {pages} {4280} (\bibinfo {year}
  {2004})}\BibitemShut {NoStop}%
\bibitem [{\citenamefont {Cao}\ \emph {et~al.}(2005)\citenamefont {Cao},
  \citenamefont {Wang},\ and\ \citenamefont {Dai}}]{Cao2005}%
  \BibitemOpen
  \bibfield  {author} {\bibinfo {author} {\bibfnamefont {J.}~\bibnamefont
  {Cao}}, \bibinfo {author} {\bibfnamefont {Q.}~\bibnamefont {Wang}}, \ and\
  \bibinfo {author} {\bibfnamefont {H.}~\bibnamefont {Dai}},\ }\href@noop {}
  {\bibfield  {journal} {\bibinfo  {journal} {Nature materials}\ }\textbf
  {\bibinfo {volume} {4}},\ \bibinfo {pages} {745} (\bibinfo {year}
  {2005})}\BibitemShut {NoStop}%
\bibitem [{\citenamefont {Bushmaker}\ \emph
  {et~al.}(2009{\natexlab{a}})\citenamefont {Bushmaker}, \citenamefont
  {Deshpande}, \citenamefont {Hsieh}, \citenamefont {Bockrath},\ and\
  \citenamefont {Cronin}}]{Bushmaker2009a}%
  \BibitemOpen
  \bibfield  {author} {\bibinfo {author} {\bibfnamefont {A.~W.}\ \bibnamefont
  {Bushmaker}}, \bibinfo {author} {\bibfnamefont {V.~V.}\ \bibnamefont
  {Deshpande}}, \bibinfo {author} {\bibfnamefont {S.}~\bibnamefont {Hsieh}},
  \bibinfo {author} {\bibfnamefont {M.~W.}\ \bibnamefont {Bockrath}}, \ and\
  \bibinfo {author} {\bibfnamefont {S.~B.}\ \bibnamefont {Cronin}},\
  }\href@noop {} {\bibfield  {journal} {\bibinfo  {journal} {Nano letters}\
  }\textbf {\bibinfo {volume} {9}},\ \bibinfo {pages} {607} (\bibinfo {year}
  {2009}{\natexlab{a}})}\BibitemShut {NoStop}%
\bibitem [{\citenamefont {Bushmaker}\ \emph
  {et~al.}(2009{\natexlab{b}})\citenamefont {Bushmaker}, \citenamefont
  {Deshpande}, \citenamefont {Hsieh}, \citenamefont {Bockrath},\ and\
  \citenamefont {Cronin}}]{Bushmaker2009b}%
  \BibitemOpen
  \bibfield  {author} {\bibinfo {author} {\bibfnamefont {A.~W.}\ \bibnamefont
  {Bushmaker}}, \bibinfo {author} {\bibfnamefont {V.~V.}\ \bibnamefont
  {Deshpande}}, \bibinfo {author} {\bibfnamefont {S.}~\bibnamefont {Hsieh}},
  \bibinfo {author} {\bibfnamefont {M.~W.}\ \bibnamefont {Bockrath}}, \ and\
  \bibinfo {author} {\bibfnamefont {S.~B.}\ \bibnamefont {Cronin}},\
  }\href@noop {} {\bibfield  {journal} {\bibinfo  {journal} {Physical review
  letters}\ }\textbf {\bibinfo {volume} {103}},\ \bibinfo {pages} {067401}
  (\bibinfo {year} {2009}{\natexlab{b}})}\BibitemShut {NoStop}%
\bibitem [{\citenamefont {Tans}\ \emph {et~al.}(1998)\citenamefont {Tans},
  \citenamefont {Verschueren},\ and\ \citenamefont {Dekker}}]{Tans1998}%
  \BibitemOpen
  \bibfield  {author} {\bibinfo {author} {\bibfnamefont {S.~J.}\ \bibnamefont
  {Tans}}, \bibinfo {author} {\bibfnamefont {A.~R.}\ \bibnamefont
  {Verschueren}}, \ and\ \bibinfo {author} {\bibfnamefont {C.}~\bibnamefont
  {Dekker}},\ }\href@noop {} {\bibfield  {journal} {\bibinfo  {journal}
  {Nature}\ }\textbf {\bibinfo {volume} {393}},\ \bibinfo {pages} {49}
  (\bibinfo {year} {1998})}\BibitemShut {NoStop}%
\bibitem [{\citenamefont {Liang}\ \emph {et~al.}(2001)\citenamefont {Liang},
  \citenamefont {Bockrath}, \citenamefont {Bozovic}, \citenamefont {Hafner},
  \citenamefont {Tinkham},\ and\ \citenamefont {Park}}]{Liang2001}%
  \BibitemOpen
  \bibfield  {author} {\bibinfo {author} {\bibfnamefont {W.}~\bibnamefont
  {Liang}}, \bibinfo {author} {\bibfnamefont {M.}~\bibnamefont {Bockrath}},
  \bibinfo {author} {\bibfnamefont {D.}~\bibnamefont {Bozovic}}, \bibinfo
  {author} {\bibfnamefont {J.~H.}\ \bibnamefont {Hafner}}, \bibinfo {author}
  {\bibfnamefont {M.}~\bibnamefont {Tinkham}}, \ and\ \bibinfo {author}
  {\bibfnamefont {H.}~\bibnamefont {Park}},\ }\href@noop {} {\bibfield
  {journal} {\bibinfo  {journal} {Nature}\ }\textbf {\bibinfo {volume} {411}},\
  \bibinfo {pages} {665} (\bibinfo {year} {2001})}\BibitemShut {NoStop}%
\bibitem [{\citenamefont {Livermore}\ \emph {et~al.}(1996)\citenamefont
  {Livermore}, \citenamefont {Crouch}, \citenamefont {Westervelt},
  \citenamefont {Campman},\ and\ \citenamefont {Gossard}}]{Livermore1996}%
  \BibitemOpen
  \bibfield  {author} {\bibinfo {author} {\bibfnamefont {C.}~\bibnamefont
  {Livermore}}, \bibinfo {author} {\bibfnamefont {C.}~\bibnamefont {Crouch}},
  \bibinfo {author} {\bibfnamefont {R.}~\bibnamefont {Westervelt}}, \bibinfo
  {author} {\bibfnamefont {K.}~\bibnamefont {Campman}}, \ and\ \bibinfo
  {author} {\bibfnamefont {A.}~\bibnamefont {Gossard}},\ }\href@noop {}
  {\bibfield  {journal} {\bibinfo  {journal} {Science}\ }\textbf {\bibinfo
  {volume} {274}},\ \bibinfo {pages} {1332} (\bibinfo {year}
  {1996})}\BibitemShut {NoStop}%
\bibitem [{\citenamefont {Katsnelson}\ \emph {et~al.}(2006)\citenamefont
  {Katsnelson}, \citenamefont {Novoselov},\ and\ \citenamefont
  {Geim}}]{Katsnelson2006}%
  \BibitemOpen
  \bibfield  {author} {\bibinfo {author} {\bibfnamefont {M.}~\bibnamefont
  {Katsnelson}}, \bibinfo {author} {\bibfnamefont {K.}~\bibnamefont
  {Novoselov}}, \ and\ \bibinfo {author} {\bibfnamefont {A.}~\bibnamefont
  {Geim}},\ }\href@noop {} {\bibfield  {journal} {\bibinfo  {journal} {Nature
  physics}\ }\textbf {\bibinfo {volume} {2}},\ \bibinfo {pages} {620} (\bibinfo
  {year} {2006})}\BibitemShut {NoStop}%
\end{thebibliography}%

\end{document}